\documentclass[twoside, twocolumn,
 floats,
 amsmath,amssymb,
 aps,
 longbibliography,
]{revtex4-1}

\usepackage[english]{babel}
\usepackage[paperwidth=210mm,paperheight=297mm,centering,hmargin=2cm,vmargin=2.5cm]{geometry}
\usepackage{graphicx}
\usepackage{diagbox}
\usepackage[colorlinks=true, allcolors=blue]{hyperref}
\usepackage{changes}
\usetikzlibrary{patterns}

\makeatletter
\renewcommand*\env@matrix[1][c]{\hskip -\arraycolsep
  \let\@ifnextchar\new@ifnextchar
  \array{*\c@MaxMatrixCols #1}}
\makeatother

\begin{document}

% \preprint{???}

\title{Artificial topological insulator realized in a two-terminal Josephson junction with Rashba spin-orbit interaction}%

\author{Luka Medic}
 \email{luka.medic@ijs.si}
 \author{Anton Ram\v{s}ak}%
\author{Toma\v{z} Rejec}%
\affiliation{Jo\v{z}ef Stefan Institute, Jamova 39, SI-1000 Ljubljana, Slovenia}
\affiliation{Faculty of Mathematics and Physics, University of Ljubljana, Jadranska	19, SI-1000 Ljubljana, Slovenia}
 
% \collaboration{MUSO Collaboration}%\noaffiliation

\date{\today}% It is always \today, today,

\begin{abstract}
    We study a two-terminal Josephson junction with conventional superconductors and a normal region with Rashba spin-orbit interaction, characterized by two Aharonov-Casher (AC) fluxes. When the superconducting phase difference equals $\pi$, the Andreev subgap spectrum may host zero-energy Weyl singularities associated with a vanishing normal-state reflection eigenvalue. With one of the AC fluxes playing the role of a quasimomentum, the junction can be viewed as an artificial one-dimensional chiral topological insulator. Its topological phase can be tuned by crossing a Weyl singularity by means of varying the remaining AC flux. By associating an additional component of the quasimomentum with the superconducting phase difference, an artificial Chern insulator is realized.
\end{abstract}

\maketitle

%\tableofcontents

% \vspace*{0pt}

% \section{Introduction}

\textit{Introduction.}---The exploration of topology in condensed matter physics has flourished in recent decades \citep{chiu2016classification, hasan2010colloquium, qi2011topological}. Notably, Riwar et al. \citep{riwar2016multi} introduced a novel class of topological systems by demonstrating that combining topologically trivial superconducting leads to form a multi-terminal Josephson junction may result in a topologically non-trivial system, an artificial topological material, where independent superconducting phase differences act as quasimomenta in a synthetic Brillouin zone (BZ). Central to their proposal were superconducting-phase-dependent Andreev bound states (ABSs), which could undergo a topological phase transition, leading to a change in the topological invariant, the Chern number. Consequently, the system exhibits quantization of the transconductance, which represents the current response in one terminal to the voltage applied to another. Eriksson et al. \citep{eriksson2017topological} numerically analyzed the same system and confirmed the quantization of the transconductance. Research was also done on three-terminal junctions in the presence of magnetic flux through the normal region \citep{meyer2017nontrivial, xie2017topological}. In these systems, the Aharonov-Bohm phase associated with the magnetic flux serves as the control parameter. Although experimental studies on multi-terminal Josephson junctions have been conducted \citep{pfeffer2014subgap, strambini2016omega, cohen2018nonlocal, draelos2019supercurrent, graziano2020transport, pankratova2020multiterminal}, none have yet demonstrated the quantization of the transconductance.

In this Letter, similar to Ref. \citep{meyer2017nontrivial}, we introduce a new control phase. Instead of the magnetic flux and the associated Aharonov-Bohm phase, we study systems exhibiting the spin-dependent Aharonov-Casher (AC) effect in the presence of the electric field \citep{aharonov1984topological, cimmino1989observation, mathur1992quantum, konig2006direct, bergsten2006experimental, liu2009control}. In quasi-1D quantum wires, the AC phase can be induced by the Rashba spin-orbit coupling \citep{rashba1959symmetry, bychkov1984properties, bychkov1984oscillatory, bihlmayer2015focus}, where electrons with opposite spins acquire opposite phases \citep{manchon2015new, tomaszewski2018aharonov1, tomaszewski2018aharonov2}. For an InGaAs-based two-dimensional electron gas, the Rashba coupling, which can be controlled by a gate voltage, typically takes values in the range of $(0.5-2.0) \times 10^{-11}$ eVm \citep{grundler2000large, citro2006pumping, wu2007analysis}. In Rashba-gate-controlled rings \citep{citro2006pumping, wu2007analysis, romeo2008quantum, tu2014real}, this corresponds to values of the AC flux between $0.3\pi$ and $3\pi$.

Studies on the effects of spin-orbit interaction in multi-terminal Josephson junctions have already been conducted \citep{beri2008splitting, yokoyama2015singularities}. These investigations examined the ABS spectrum, including the impact of spin-orbit interaction on Weyl nodes and superconducting gap edge touchings. However, the consideration of spin-orbit phenomena as a new possibility for the quasimomentum of a synthetic BZ has not been explored in these works.

\begin{figure}
    \begin{tikzpicture}
        % Normal region
        \draw[fill=orange!50] (-3,-0.7) rectangle (3,0.7);
        \draw[rounded corners=2mm, fill=white] (-3.5,-0.6) rectangle (-0.05,0.6);
        \draw[rounded corners=2mm, fill=white] ( 0.05,-0.6) rectangle ( 3.5,0.6);
    
        % Leads
            % left
            \draw[rounded corners=0mm, fill=gray!30, draw=none]  (-4,-1) rectangle (-2.5,1);
            \draw[rounded corners=0mm, fill=white] (-4,-1)-++(1.5,0);
            \draw[rounded corners=0mm, fill=white] (-4, 1)-++(1.5,0);
            \draw[rounded corners=0mm, fill=white] (-2.5,-1)-++(0,2);
            % right
            \draw[rounded corners=0mm, fill=gray!30, draw=none]  ( 2.5,-1) rectangle ( 4,1);
            \draw[rounded corners=0mm, fill=white] ( 4,-1)-++(-1.5,0);
            \draw[rounded corners=0mm, fill=white] ( 4, 1)-++(-1.5,0);
            \draw[rounded corners=0mm, fill=white] ( 2.5,-1)-++(0,2);
    
        % % Annotate
        \node at (-3.3,0) {$\phi/2$};
        \node at ( 3.3,0) {$-\phi/2$};
    
        % % Rashba
        \draw[pattern=north east lines, draw=none] (-1.35-0.5,-0.65-0.2) rectangle (-1.35+0.5,-0.65+0.2);
        \draw[pattern=north east lines, draw=none] ( 1.35-0.5,-0.65-0.2) rectangle ( 1.35+0.5,-0.65+0.2);
        
        \node at (-1.35,-0.2) {$\alpha_1$};
        \node at ( 1.35,-0.2) {$\alpha_2$};
        
    \end{tikzpicture}
    \caption{SNS Josephson junction with a superconducting phase difference $\phi$ and a normal region (orange) consisting of inter-connected quantum wires, forming two rings (genus 2). Each ring can be assigned an independent AC flux ($\alpha_{1,2}$), controllable via Rashba coupling using gate electrodes (hatched).}
    \label{fig:sketch}
\end{figure}
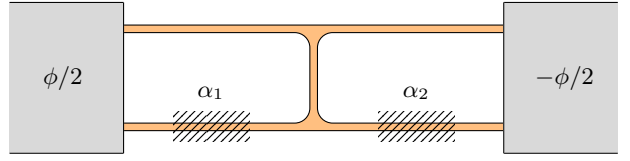

Here, we present a theoretical analysis focused on the exploration of low-dimensional, topologically non-trivial artificial materials utilizing the Rashba interaction within the normal region. The minimal model consists of a two-terminal Josephson junction with a normal region composed of two rings, as depicted in Fig. \ref{fig:sketch}, each permeated by an AC flux. Our investigation concludes that both the winding number and the Chern number can be identified in this system. For the Chern number, the system's synthetic BZ consists of an AC flux and the superconducting phase difference, whereas for the winding number, the superconducting phase difference is fixed to $\pi$.

The Letter is organized to first analyze the symmetries of a general multi-terminal system with multiple AC fluxes in the normal region and obtain the topological classification for this family of systems. Subsequently, we restrict ourselves to the analysis of systems with only two leads and two AC fluxes. The winding number is computed, and with the help of the low-energy Hamiltonian at the Weyl node, its connection to the Chern number is established. Utilizing the system's symmetries, we conclude that for each spin sector, Weyl nodes come in quartets with pairs of opposite topological charges.

% \section{Symmetries}
\textit{Symmetries.}---We consider a general case of a multi-terminal Josephson junction in the presence of AC fluxes in the normal region. The parameters of our model encompass a vector $\vec{\phi}$ containing the independent superconducting phase differences, and a vector $\vec{\alpha}$ describing the AC fluxes through the normal region. Specifically, there is one AC flux for each ring (refer to Fig. \ref{fig:sketch}).

We begin with the electron Hamiltonian, which describes the system in its normal state and has the following structure \citep{citro2006pumping, wu2007analysis}:
\begin{equation}
    H(\vec{\alpha}) = \begin{bmatrix}
        H_{\uparrow}(\vec{\alpha}) & 0 \\
        0 & H_{\downarrow}(\vec{\alpha}) \\
    \end{bmatrix}.
\end{equation}
Since electrons with opposite spins acquire opposite AC phases, the non-zero blocks satisfy the relations $H_{\downarrow}(\vec{\alpha}) = H_{\uparrow}(-\vec{\alpha}) = H_{\uparrow}(\vec{\alpha})^*$. Consequently, $H$ exhibits symmetry under conjugation and time-reversal symmetry (TRS):
\begin{subequations}
    \label{eq:sym:conjugationSym_TRS:H:electron}
    \begin{align}
        \label{eq:sym:conjugationSym:H:electron}
        H(\vec{\alpha})^* = H(-\vec{\alpha}), \\
        \label{eq:sym:TRS:H:electron}
        T H(\vec{\alpha}) T^{-1} = H(\vec{\alpha}),
    \end{align}
\end{subequations}
where $T = i \sigma_y \mathcal{K}$; here, $\sigma_y$ is the Pauli matrix operating on the spin degree of freedom, and $\mathcal{K}$ is the conjugation operator.

This allows us to represent the Bogoliubov-de Gennes (BdG) Hamiltonian, which incorporates the description of superconductivity, as  
\begin{equation}
\label{eq:H:BdG}
    H_{\textrm{BdG}}(\vec{\phi}, \vec{\alpha}) = \begin{bmatrix}
        H(\vec{\alpha}) & \Delta(\vec{\phi}) \\
        \Delta(\vec{\phi})^\dagger & -T H(\vec{\alpha}) T^{-1} \\
    \end{bmatrix},
\end{equation}
where $\Delta(\vec{\phi})$ is a diagonal matrix for the s-wave pairing. In this form of the BdG Hamiltonian, the intrinsic particle-hole symmetry (PHS) is expressed as
\begin{equation}
    \label{eq:sym:intrinsicPHS}
    \tau_y \sigma_y H_{\textrm{BdG}}^*(\vec{\alpha}, \vec{\phi}) \sigma_y \tau_y = - H_{\textrm{BdG}}(\vec{\alpha}, \vec{\phi}),
\end{equation}
where $\tau_y$ is the Pauli matrix acting in the Nambu space. Additionally, from Eq. (\ref{eq:sym:conjugationSym_TRS:H:electron}) and the relations $T \Delta(\vec{\phi}) T^{-1} = \Delta(\vec{\phi})^* = \Delta(-\vec{\phi})$, it follows that the BdG Hamiltonian also exhibits symmetry under conjugation and TRS:
\begin{subequations}
    \label{eq:sym:conjugation_TRS}
    \begin{align}
        \label{eq:sym:conjugation}
        H_{\textrm{BdG}}^*(\vec{\alpha}, \vec{\phi}) &=  H_{\textrm{BdG}}(-\vec{\alpha}, -\vec{\phi}), \\
        \label{eq:sym:TRS}
        \sigma_y H_{\textrm{BdG}}^*(\vec{\alpha}, \vec{\phi}) \sigma_y &=  H_{\textrm{BdG}}(\vec{\alpha}, -\vec{\phi}).
    \end{align}
\end{subequations}

The BdG Hamiltonian is symmetric under $\sigma_z$, i.e. $\sigma_z H_{\textrm{BdG}}(\vec{\alpha}, \vec{\phi}) \sigma_z = H_{\textrm{BdG}}(\vec{\alpha}, \vec{\phi})$,
so the Hilbert space splits into two spin sectors. Hence, our focus lies on the classification within each sector. Combining Eqs. (\ref{eq:sym:intrinsicPHS}) and (\ref{eq:sym:conjugation_TRS}), we obtain two symmetries that do not mix the spin sectors:
\begin{subequations}
    \begin{eqnarray}
    \label{eq:sym:sector:PHS}
        \tau_y H_{\textrm{BdG}}^*(\vec{\alpha}, \vec{\phi}) \tau_y = - H_{\textrm{BdG}}(-\vec{\alpha}, \vec{\phi}), \\
        \label{eq:sym:sector:chiral}
        \tau_y H_{\textrm{BdG}}(\vec{\alpha}, \vec{\phi}) \tau_y = - H_{\textrm{BdG}}(\vec{\alpha}, -\vec{\phi}),
    \end{eqnarray}
\end{subequations}
which can be regarded as PHS and chiral symmetry (CS) within each sector. Note that, as a combination of TRS and the intrinsic PHS, chiral symmetry is generally preserved even in more complex Rashba interaction models.

% \section{Topological classification}
\textit{Topological classification.}---To transition a system between different topological phases, it is necessary to introduce at least one control parameter. Here, we focus on two scenarios: using either a single superconducting phase difference or a single AC flux as the control parameter. Selecting a superconducting phase difference as the control parameter breaks the CS described in Eq. (\ref{eq:sym:sector:chiral}) while preserving the PHS described in Eq. (\ref{eq:sym:sector:PHS}). Conversely, choosing an AC flux as the control parameter breaks the PHS while leaving the CS unaffected. Referring to Ref. \citep{shiozaki2014topology}, we have constructed topological classification tables for both scenarios (see Tables \ref{tab:topological_classification:1} and \ref{tab:topological_classification:2}); for specifics, refer to Supplemental Materials (SM) S1 \citep{medic2024supplemental}. In the tables, $d_\phi$ and $d_\alpha$ denote the dimensionalities of phases composing the synthetic BZ, culminating in its total dimensionality $d_{\textrm{BZ}} = d_\phi + d_\alpha$.

In Table \ref{tab:topological_classification:1}, the case [$d_\phi = 2, d_\alpha = 0$], having a topological invariant $\mathbb{Z}$, corresponds to the system with a non-trivial Chern number investigated in Ref. \citep{riwar2016multi}. Specifically, the minimal model constitutes a four-terminal Josephson junction featuring three independent superconducting phases. Among these phases, two contribute to defining the synthetic BZ, while the remaining one functions as the control phase.

From here on, we restrict our analysis to systems with two superconducting leads only. In Table \ref{tab:topological_classification:1}, another consideration for identifying topologically nontrivial phases in low-dimensional systems is the option [$d_\phi = 0, d_\alpha = 2$]. In this case, the control parameter is the superconducting phase difference $\phi$. Yet,  as we will delve into in the subsequent sections, all the topological charges are situated at $\phi = \pi$ with a total charge of 0. Consequently, the behavior manifests as trivial.

\begin{table}
\begin{tabular}{c|cccccccc}
\diagbox[height=0.6cm,width=0.8cm]{$d_\alpha$}{$d_\phi$}  & 0 & 1 & 2 & 3 & 4 & 5 & 6 & 7 \\
\hline
0 & 0 & 0 & $\mathbb{Z}$ & $\mathbb{Z}_2$ & $\mathbb{Z}_2$ & 0 & $2\mathbb{Z}$ & 0 \\
1 & 0 & 0 & 0 & $\mathbb{Z}$ & $\mathbb{Z}_2$ & $\mathbb{Z}_2$ & 0 & $2\mathbb{Z}$ \\
2 & $2\mathbb{Z}$ & 0 & 0 & 0 & $\mathbb{Z}$ & $\mathbb{Z}_2$ & $\mathbb{Z}_2$ & 0 \\
3 & 0 & $2\mathbb{Z}$ & 0 & 0 & 0 & $\mathbb{Z}$ & $\mathbb{Z}_2$ & $\mathbb{Z}_2$ \\
4 & $\mathbb{Z}_2$ & 0 & $2\mathbb{Z}$ & 0 & 0 & 0 & $\mathbb{Z}$ & $\mathbb{Z}_2$ \\
5 & $\mathbb{Z}_2$ & $\mathbb{Z}_2$ & 0 & $2\mathbb{Z}$ & 0 & 0 & 0 & $\mathbb{Z}$ \\
6 & $\mathbb{Z}$ & $\mathbb{Z}_2$ & $\mathbb{Z}_2$ & 0 & $2\mathbb{Z}$ & 0 & 0 & 0 \\
7 & 0 & $\mathbb{Z}$ & $\mathbb{Z}_2$ & $\mathbb{Z}_2$ & 0 & $2\mathbb{Z}$ & 0 & 0 \\
\end{tabular}
\caption{Classification table for the case where the control parameter is a superconducting phase difference, resulting in the exclusive presence of PHS [Eq. (\ref{eq:sym:sector:PHS})] within each spin sector. Bott periodicity implies a periodic structure in $d_{\phi}$ and $d_{\alpha}$ with a period of $8$.}
\label{tab:topological_classification:1}
\end{table}

Selecting one of the AC fluxes as the control parameter, Table \ref{tab:topological_classification:2} reveals that the lowest dimensional nontrivial cases are $\mathbb{Z}$ and $\mathbb{Z} \oplus \mathbb{Z}$ for [$d_\phi = 0, d_\alpha = 1$] and [$d_\phi = 1, d_\alpha = 1$], respectively. Here, $\mathbb{Z}$ represents the winding number, while $\mathbb{Z} \oplus \mathbb{Z}$ corresponds to the Chern number and the mirror winding number \citep{chiu2013classification, shiozaki2014topology, zhang2014anomalous}. In the former case, the presence of CS requires the superconducting phase difference to be either $0$ or $\pi$. Additionally, to allow for gap closures and enable topological phase transitions, we set $\phi = \pi$. Conversely, in the latter case, this constraint is lifted and $\phi$ is treated as a quasimomentum. For the remainder of the paper, our focus will be on these two cases, i.e. a two-terminal Josephson junction featuring two AC fluxes in the normal region.

\begin{table}
\begin{tabular}{c|cc}
\diagbox[height=0.6cm,width=0.8cm]{$d_\alpha$}{$d_\phi$}  & even & odd \\
\hline
even & 0 & 0 \\
odd & $\mathbb{Z}$ & $\mathbb{Z} \oplus \mathbb{Z}$ \\
\end{tabular}
\caption{Classification table for the case where an AC flux is used as the control parameter, maintaining CS [Eq. \eqref{eq:sym:sector:chiral}] within each spin sector.}
\label{tab:topological_classification:2}
\end{table}

% \section{Effective Hamiltonian}
% \label{section:effectiveHamiltonian}

\textit{Effective Hamiltonian.}---We begin with the expression determining the Andreev bound states with energy $E$ ($|E| < |\Delta|$) within the sector corresponding to electrons (holes) with spin $\uparrow$ ($\downarrow$) \citep{beenakker1991universal, nazarov_blanter_2009, yokoyama2015singularities}:
\begin{equation}
\label{eq:scatteringABS}
    \begin{bmatrix}
        0 & S_\uparrow(\vec{\alpha})\, e^{i \hat{\phi}} \\
        S_\downarrow^*\!(\vec{\alpha})\, e^{-i \hat{\phi}} & 0 \\
    \end{bmatrix} \psi = e^{i \chi} \psi,
\end{equation}
where $\chi \equiv \arccos(E/|\Delta|)$, with leads having the same gap $|\Delta|$. The components of $\psi = \left[\psi_e, \, \psi_h \right]^T$ are the amplitudes of the outgoing waves for electrons and holes. The matrix $\hat{\phi}= \textrm{diag}(\phi/2, -\phi/2) \otimes \mathbb{I}_{\scriptscriptstyle N}$ is diagonal, where $\phi \in \left[0, 2\pi\right)$ denotes the superconducting phase difference, and $\mathbb{I}_{\scriptscriptstyle N}$ signifies the identity matrix with dimension $N$, indicating the number of modes in each lead. (We assume for simplicity that the number of modes in both leads is the same.) $S_{\uparrow}$ ($S_{\downarrow}^*$) represents the normal-state scattering matrix for electrons (holes) with spin $\uparrow$ ($\downarrow$); it is worth noting that $S_{\downarrow}(\vec{\alpha}) = S_{\uparrow}^T\!(\vec{\alpha})$. In the following, we will focus on the short junction limit, described by an energy-independent scattering matrix, and use the abbreviation $S = S_{\uparrow}(\vec{\alpha})$. A longer scattering region, characterized by an energy-dependent matrix $S(E)$, would lead to additional ABSs at finite energy. Nevertheless, the existence of Weyl nodes and their topological properties at zero energy depend solely on $S(0)$.

From Eq.~(\ref{eq:scatteringABS}), the equation for the electron part \footnote{In Supplemental Materials S2 \citep{medic2024supplemental}, we present an alternative approach that considers both the electron and hole parts simultaneously. We demonstrate that the conclusions remain consistent with those obtained using the approach presented in this paper, which considered only the electron part when computing the effective Hamiltonian and the topological invariants.} can be obtained \citep{beenakker1991universal, riwar2016multi}:
\begin{equation}
    \label{eq:scatteringABSelectronPart}
    A \psi_e = e^{2i \chi} \psi_e,
\end{equation}
where $A=S e^{i \hat{\phi}} S^\dagger e^{-i \hat{\phi}}$ is unitary. Similar to the Joukowsky transform \citep{van2014single}, we decompose $A=A_1+i A_2$ into its hermitian and antihermitian component, where $A_1=\frac{1}{2}\left(A+A^\dagger\right)$ and $A_2=\frac{1}{2i}\left(A-A^\dagger\right)$ are both hermitian matrices. Separating the two parts of Eq.~(\ref{eq:scatteringABSelectronPart}) yields a system of equations:
\begin{subequations}
    \begin{align}
        \label{eq:A1_system}
        \left(\frac{\mathbb{I}_{\scriptscriptstyle N}-A_1}{2}\right)^{1/2} \psi_e = \sin \chi \,\,\psi_e, \\
        \label{eq:A2_system}
        A_2 \psi_e = 2\cos\chi \sin\chi \,\,\psi_e,
    \end{align}
\end{subequations}
where
\begin{subequations}
    \begin{align}
    \mathbb{I}_{\scriptscriptstyle N} - A_1 &= 2\sin^2 \frac{\phi}{2} \begin{bmatrix}
        t' t'^\dagger & 0 \\
        0 & t t^\dagger \\
    \end{bmatrix}, \\
    A_2 &= 2 \sin \frac{\phi}{2} \begin{bmatrix}
        -t' t'^\dagger \cos\frac{\phi}{2} & r t^\dagger e^{i \frac{\phi}{2}} \\
        t r^\dagger e^{-i \frac{\phi}{2}} & t t^\dagger \cos\frac{\phi}{2} \\
    \end{bmatrix}.
    \end{align}
\end{subequations}
Here, $r$ ($r'$) and $t$ ($t'$) are $N \times N$ reflection and transmission matrices at the left (right) lead, respectively, and $(\mathbb{I}_{\scriptscriptstyle N}-A_1)$ is positive semi-definite. In Eq. (\ref{eq:A1_system}), we took into account the constraint $\textrm{Im}(e^{i \chi}) \geq 0$ \citep{beenakker1991universal}, i.e. $\chi \in \left[0, \pi\right]$. Assuming that $(\mathbb{I}_{\scriptscriptstyle N} - A_1)$ and $A_2$ are non-singular (requiring $\phi \neq 0$), we can combine Eqs. (\ref{eq:A1_system}) and (\ref{eq:A2_system}) to derive the effective (electron) Hamiltonian acting in the ABS subspace, $H_e \psi_e = E \psi_e$, given by
\begin{equation}
    \label{eq:effectiveElectronHamiltonian}
    \!H_e = |\Delta|\begin{bmatrix}
        -(t' t'^\dagger)^{1/2} \cos\frac{\phi}{2} & r t^\dagger (t t^\dagger)^{-1/2} e^{i \phi/2} \\
        (t t^\dagger)^{-1/2} t r^\dagger e^{-i \phi/2} & (t t^\dagger)^{1/2} \cos\frac{\phi}{2} \\
    \end{bmatrix}.
\end{equation}
The eigenvalues of $H_e$ come in pairs $E_n^{\pm} = \pm|\Delta| \sqrt{1- T_n \sin^2(\phi/2)}$, where $T_n$ are eigenvalues of $t t^\dagger$ \citep{beenakker1991universal}. Thus, the gap closes when $\phi = \pi$ and $T_n = 1$. In the limit $\phi \to 0$, ABSs approach the continuum at $\pm |\Delta|$.

% \section{Chiral symmetry}
\textit{Chiral symmetry.}---The chiral symmetry [Eq. (\ref{eq:sym:sector:chiral})] implies that eigenfunctions appear in pairs with opposite energies, satisfying $\Psi(\vec{\alpha}, -\vec{\phi}) = i \tau_y \Psi(\vec{\alpha}, \vec{\phi})$. This results in CS effectively swapping the roles of incoming and outgoing waves. Additionally, incoming waves are related to outgoing waves through Andreev reflection \citep{beenakker1991universal}. Consequently, under chiral symmetry, the components of outgoing waves $(\psi_e, \psi_h)$ transform to $(e^{-i\hat{\phi}-i\chi} \psi_e, -e^{i\hat{\phi}-i\chi} \psi_h)$.

Focusing on the electron component, the effective Hamiltonian from the previous section satisfies $e^{i\hat{\phi}} H_e(\vec{\alpha}, 2\pi-\phi) e^{-i\hat{\phi}} = - H_e(\vec{\alpha}, \phi)$. For $\phi = \pi$, this reduces to $\Gamma_\pi H_e \Gamma_\pi^\dagger = -H_e$, where $\Gamma_\pi = \nu_z \otimes \mathbb{I}_{\scriptscriptstyle N}$ and $\nu_z$ is the Pauli matrix acting in the space of left and right outgoing waves. This is unsurprising, as $H_e$ becomes block off-diagonal at $\phi = \pi$.

% \section{Topological invariants and the low-energy Hamiltonian}
\textit{Topological invariants and the low-energy Hamiltonian.}---We proceed by determining the topological invariant at $\phi=\pi$. From the topological classification, we already know that the appropriate invariant is the winding number \citep{shiozaki2014topology, maffei2018topological}. However, because the winding number is gauge-dependent, we will consider the differences between winding numbers at different values of the control AC flux. These differences correspond to contractible (i.e., null-homotopic) loops in the space of AC fluxes. For such loops, we derive:
\begin{equation}
    W_\pi = \frac{1}{2\pi i} \oint_{\mathcal{C}_\alpha} \textrm{d} \log \left( \textrm{det}\, h \right) = \frac{1}{2\pi i} \oint_{\mathcal{C}_\alpha} \textrm{d} \log \left( \textrm{det}\, r \right),
\end{equation}
where $\mathcal{C}_\alpha$ is an arbitrary contractible loop, and $h = i r t^\dagger (t t^\dagger)^{-1/2}$ is the off-diagonal block of $H_e$. In the last step, we assumed that $\mathcal{C}_\alpha$ does not cross or enclose any gap edge touchings. Consequently, only the reflection matrix $r$ contributes to $W_\pi$.

Next, let us focus on the low-energy limit, particularly on the band touching at a Weyl node located at $\vec{x}_{\scriptscriptstyle W} = [\alpha_{1{\scriptscriptstyle W}}, \alpha_{2{\scriptscriptstyle W}}, \pi]^T$. Denoting the two chiral states at the Weyl node with positive and negative chirality as $|a_e^+ \rangle$ and $|a_e^- \rangle$, respectively (where $\nu_z | a_e^\pm \rangle = \pm | a_e^\pm \rangle$), we can express the low-energy Hamiltonian as (details in SM S2 \citep{medic2024supplemental})
\begin{equation}
    H_{\scriptscriptstyle W} = \delta \vec{x} \cdot M_3 \vec{\Sigma}.
\end{equation}
Here, $\delta \vec{x} = [\delta \alpha_1, \delta \alpha_2, \delta \phi]^T$ represents the displacement from the Weyl node, $\vec{\Sigma}$ is a vector of Pauli matrices acting in the space of chiral states $|a_e^\pm \rangle$, and
\begin{equation}
    \label{eq:lowEnergyLimit:M3}
    M_3 \!=\!\! 
    \left[ \begin{array}{ *{4}{c} }
    & & 0 \\
    \multicolumn{2}{c}
      {\raisebox{\dimexpr\normalbaselineskip+.7\ht\strutbox-1.2\height}[0pt][0pt]{\scalebox{1.5}{$M_2$}}} & 0 \\
    0 & 0 & \frac{1}{2}
    \end{array} \right]\!\!,\, M_2 \!=\!\! \begin{bmatrix}
    \partial_{\alpha_1} {\rm Re}(\zeta) & -\partial_{\alpha_1} {\rm Im}(\zeta) \\
    \partial_{\alpha_2} {\rm Re}(\zeta) & -\partial_{\alpha_2} {\rm Im}(\zeta) \end{bmatrix}_{\vec{x}= \vec{x}_{\scriptscriptstyle W}}
\end{equation}
where $\zeta(\vec{x}) = \langle a_e^+ | H_e(\vec{x}) | a_e^- \rangle$, and $M_2$ is evaluated at the Weyl node. Thus, the low-energy dispersion at the Weyl node is conical, given by $E^\pm = \pm\sqrt{\left|\delta\vec{\alpha} \cdot \nabla_{\vec{\alpha}} \zeta\right|^2 + (\delta\phi/2)^2}$.

From Eq. (\ref{eq:lowEnergyLimit:M3}), topological charges associated with winding and Chern numbers can be obtained as $W_{\scriptscriptstyle W} = \textrm{sgn} \left(\textrm{det}(M_2) \right)$ and $C_{\scriptscriptstyle W} = \textrm{sgn} \left(\textrm{det}(M_3) \right)$, respectively. It is clear that the two charges are equivalent, and we henceforth denote them as $q_{\scriptscriptstyle W}$.

Following the topological classification of the BdG Hamiltonian presented in Table \ref{tab:topological_classification:2}, the remaining invariant is the mirror winding number, defined as \citep{chiu2013classification}
\begin{equation}
    W_{M\mathbb{Z}} = \textrm{sgn}\left(W_0 - W_\pi\right) \left(|W_0| - |W_\pi|\right),
\end{equation}
where $W_0$ is the winding number evaluated in the plane $\phi = 0$. Although the winding number is well-defined for the entire BdG Hamiltonian, our approach involving scattering matrices and ABSs, which merge with the continuum, does not permit us to obtain $W_0$ in our present formulation. We leave the calculation of the mirror winding number of the BdG Hamiltonian for further study.

Applying conjugation symmetry, Eq. (\ref{eq:sym:conjugation}), at the conjugation-invariant phase $\phi=\pi$ \footnote{Here, the conjugation symmetry, Eq. (\ref{eq:sym:conjugation}), mimics TRS within each distinct sector, where $\phi=\pi$ is analogous to time-reversal invariant momentum (TRIM).}, leads us to the conclusion that Weyl nodes come in pairs at $\pm \vec{x}_{\scriptscriptstyle W}$ with equal topological charges, $q_{\scriptscriptstyle W}(\vec{x}_{\scriptscriptstyle W}) = q_{\scriptscriptstyle W}(-\vec{x}_{\scriptscriptstyle W})$. Given this equivalence of topological charges and considering the Nielsen-Ninomiya theorem, which necessitates total charge cancellation \citep{nielsen1981absence1, nielsen1981absence2, murakami2007phase}, we infer the existence of quartets of Weyl nodes within each spin sector. Each quartet comprises of two Weyl nodes with positive topological charge and two with negative topological charge.

Considering both sectors, the TRS [Eq. (\ref{eq:sym:TRS})] implies that $q_{\scriptscriptstyle W}^{\downarrow}(\vec{x}_{\scriptscriptstyle W}) = q_{\scriptscriptstyle W}^{\uparrow}(-\vec{x}_{\scriptscriptstyle W}) = q_{\scriptscriptstyle W}^{\uparrow}(\vec{x}_{\scriptscriptstyle W})$, resulting in both sectors having identical topological charges at $\vec{x}_{\scriptscriptstyle W}$. Consequently, eigenstates at the Weyl nodes are topologically protected and can be gapped only by the annihilation of opposite charges. The total charge, combining charges for both sectors, amounts to twice the topological charge for each individual sector.

% \section{Discussion}

\textit{Discussion.}---To summarize, we provided a topological classification of multi-terminal Josephson junctions with AC fluxes in the normal region. In contrast to previous proposals for realizing Josephson junctions as topological matter \citep{riwar2016multi, meyer2017nontrivial, xie2017topological}, we showed that topologically non-trivial regimes can arise even in Josephson junctions with just two terminals. Specifically, for two-terminal Josephson junctions with the superconducting phase difference equal to $\pi$, we demonstrated that when no gap edge touchings are enclosed by $\mathcal{C}_\alpha$, the computation of winding number differences simplifies to the topological properties of the reflection matrix $r$. We confirmed that the dispersion near zero-energy singularities is conical, identifying these singularities as Weyl nodes. Additionally, we established the equivalence between the topological charges associated with the winding number and the Chern number. Finally, we showed that the Weyl nodes of the two spin sectors coincide and carry the same topological charges.

In our approach, we derived ABS spectra using scattering matrices. However, by neglecting the energy dependence of these matrices, we encountered ABS spectra with gap edge touchings. Incorporating energy dependence in the scattering matrix could introduce finite level repulsion between the highest ABS and the continuum, thereby restoring the adiabaticity of the ABSs, as discussed in Ref. \citep{riwar2016multi}. Furthermore, as highlighted in the main text, another limitation of this scattering formalism and ABS-based approach is its inability to compute the mirror winding number of the BdG Hamiltonian. To address this, analytical generalizations similar to those proposed in Ref. \citep{repin2019topological} may be required.

Another challenge stems from the constrained variation range of the AC flux. To handle this limitation, a protocol should be devised that does not require quasimomentum variations exceeding $2\pi$, unlike the protocol presented in Ref. \citep{riwar2016multi}. The proposed protocol should identify a quantized physical observable related to either the winding number or the Chern number (or their associated topological charges). It is important to note that, unlike superconducting phases, which are associated with electric currents, AC fluxes are related to spin currents.

Further exploration of these ideas is a direction for future theoretical research. While we acknowledge that experimentally realizing the proposed systems remains challenging, we hope our results will inspire efforts toward the development of such devices.

\vspace*{\baselineskip}
% \section*{Acknowledgments}
We appreciate valuable discussions with Y. V. Nazarov and R. \v{Z}itko. The authors acknowledge support from the Slovenian Research and Innovation Agency (ARIS) under Contract No. P1-0044; AR was also supported by Grant J2-2514.

% \bibliographystyle{ieeetr}
% \bibliographystyle{apsrev4-1}
%\bibliography{bibliography_list}

\end{document}

% \preprint{???}

\title{Supplemental Materials\\[0.5cm]Artificial topological insulator realized in a two-terminal Josephson junction with Rashba spin-orbit interaction}%

\author{Luka Medic}
 \email{luka.medic@ijs.si}
 \author{Anton Ram\v{s}ak}%
\author{Toma\v{z} Rejec}%
\affiliation{Jo\v{z}ef Stefan Institute, Jamova 39, SI-1000 Ljubljana, Slovenia}
\affiliation{Faculty of Mathematics and Physics, University of Ljubljana, Jadranska	19, SI-1000 Ljubljana, Slovenia}
 
% \collaboration{MUSO Collaboration}%\noaffiliation

\date{\today}% It is always \today, today,

\maketitle

%\tableofcontents

% \vspace*{0pt}

% \appendix

\onecolumngrid

\section{Topological classification using K-theory}
\label{appendix:topologicalClassification}

We regard the PHS symmetry [Eq. (4a) of the main text] as an additional anti-symmetric anti-unitary symmetry within class A of the standard Altland-Zirnbauer (AZ) periodic table. Following the K-theory framework outlined in Ref. \citep{shiozaki2014topology}, we derive
\begin{align}
\label{eq:Ktheory:1}
    K_{\mathbb{C}}^A\!&\left(s\!=\!6;\,d\!=\!d_{\phi}\!+\!d_{\alpha},\, d_{\parallel}\!=\!d_{\phi},\, D\!=\!0,\, D_{\parallel}\!=\!0 \right) \nonumber \\
    &= K_{\mathbb{C}}^A\!\left(6+d_{\phi}-d_{\alpha};\,0,\,0,\, 0,\, 0 \right) \\
    &= \pi_{0}\left(\mathcal{R}_{6+d_{\phi}-d_{\alpha}}\right) \nonumber
\end{align}
where Bott periodicity implies $\mathcal{R}_p \simeq \mathcal{R}_{p+8}$, $\pi_{0}$ is the 0th homotopy ‘group’, and $\mathcal{R}$ represents the classifying space of real AZ classes. From  Eq. (\ref{eq:Ktheory:1}) the Table I of the main text follows.

Similarly, the chiral symmetry [Eq. (4b) of the main text] is regarded as an additional unitary anti-symmetry within class A. This leads to the following expression:
\begin{eqnarray}
\label{eq:Ktheory:2}
    K_{\mathbb{C}}^A\!&\left(s\!=\!0,\,t\!=\!1;\,d\!=\!d_{\phi}\!+\!d_{\alpha},\, d_{\parallel}\!=\!d_{\phi},\, D\!=\!0,\, D_{\parallel}\!=\!0 \right) \nonumber \\
    &= K_{\mathbb{C}}^A\!\left(0-d_{\phi}-d_{\alpha},\,1-d_{\phi};\,0,\,0,\, 0,\, 0 \right).
\end{eqnarray}
If $d_{\phi}$ is odd, Eq. (\ref{eq:Ktheory:2}) reduces to
\begin{equation}
    \pi_0(\mathcal{C}_{-d_{\phi}-d_{\alpha}}) \oplus \pi_0(\mathcal{C}_{-d_{\phi}-d_{\alpha}}),
\end{equation}
otherwise
\begin{equation}
    \pi_0(\mathcal{C}_{1-d_{\phi}-d_{\alpha}}),
\end{equation}
where $\mathcal{C}$ represents the classifying space of complex AZ classes, and due to Bott periodicity we have $\mathcal{C}_p \simeq \mathcal{C}_{p+2}$. Knowing that $\pi_{0}(\mathcal{C}_0) = \mathbb{Z}$ and $\pi_{0}(\mathcal{C}_1) = 0$, Table II of the main text is obtained.

\section{Alternative approach utilizing the full Nambu space}
\label{appendix:hamiltonian}

Below, we provide an alternative approach to obtain the effective ABS Hamiltonian. Instead of considering only the electron part as in the main text, we include the hole part as well. After deriving the effective Hamiltonian, we rewrite the chiral symmetry in the basis of outgoing waves. Following this, we compute the winding number and the low-energy two-level Hamiltonian. We show that this approach yields results identical to those presented in the main text, where only the electron part was considered.

\subsection{Effective Hamiltonian}

By the means of polar decomposition of the scattering matrix $S$ \citep{martin1992wave, beenakker1997random}, let us transform the Eq. (5) of the main text. Polar decomposition of the scattering matrix $S$ can be written as
\begin{equation}
\label{eq:polarDecomposition}
    S = V \tilde{S} U^\dagger
\end{equation}
where $V$ and $U$ are block diagonal unitary matrices
\begin{equation}
    V = \begin{bmatrix}
    V_1 & 0 \\
    0 & V_2 \\
    \end{bmatrix}, \quad U = \begin{bmatrix}
    U_1 & 0 \\
    0 & U_2 \\
    \end{bmatrix},
\end{equation}
and
\begin{equation}
    \tilde{S} = \begin{bmatrix}
    -i \sqrt{\mathcal{R}} & \sqrt{\mathcal{T}} \\
    \sqrt{\mathcal{T}} & -i \sqrt{\mathcal{R}} \\
    \end{bmatrix}.
\end{equation}
Here, $\mathcal{R}=\textrm{diag}(\mathcal{R}_n)$ and $\mathcal{T}=\mathbb{I}_{\scriptscriptstyle N} - \mathcal{R}$ are diagonal matrices, consisting of singular values of $r r^\dagger$ and $t t^\dagger$, respectively. Now let us write a unitary transformation that does not mix electron and hole parts of the wavefunction:
\begin{equation}
    \mathcal{U} = \begin{bmatrix}
        \mathcal{H}_{\scriptscriptstyle 2N} e^{-i \hat{\phi}/2} V^\dagger & 0 \\
        0 & \mathcal{H}_{\scriptscriptstyle 2N} e^{i \hat{\phi}/2} U^\dagger
    \end{bmatrix},
\end{equation}
where $\mathcal{H}_{\scriptscriptstyle 2N}$ represents the Hadamard matrix tensored with $\mathbb{I}_{\scriptscriptstyle N}$:
\begin{equation}
    \mathcal{H}_{\scriptscriptstyle 2N} = \frac{1}{\sqrt{2}} \begin{bmatrix}[r]
        1 & 1 \\
        1 & -1 \\
    \end{bmatrix} \otimes \mathbb{I}_{\scriptscriptstyle N}.
\end{equation}
Note that $e^{i \hat{\phi}/2} = \textrm{diag}(e^{i\phi/4}, e^{-i\phi/4}) \otimes \mathbb{I}_{\scriptscriptstyle N}$ commutes with $V$ and $U$. $\mathcal{U}$ transforms Eq. (5) of the main text to
\begin{equation}
    \begin{bmatrix}
        0 & S'\\
        (S')^* & 0
    \end{bmatrix} \psi' = e^{i\chi} \psi'
\end{equation}
where $\psi' = \mathcal{U} \psi$, and
\begin{equation}
    S' = \mathcal{H}_{\scriptscriptstyle 2N} e^{-i \hat{\phi}/2} \tilde{S} e^{i \hat{\phi}/2} \mathcal{H}_{\scriptscriptstyle 2N} = \begin{bmatrix}
        \mathcal{Z} & i \mathcal{W} \\
        -i \mathcal{W} & -\mathcal{Z}^*
    \end{bmatrix}
\end{equation}
where $\mathcal{Z} = \sqrt{\mathcal{T}} \cos(\phi/2) - i \sqrt{\mathcal{R}}$ and $\mathcal{W} = \sqrt{\mathcal{T}} \sin(\phi/2)$. Solving the eigenproblem for the electron part of the wavefunction
\begin{equation}
    A' \psi'_e = e^{2i\chi} \psi'_e,
\end{equation}
where
\begin{equation}
    A' = S' (S')^* = \begin{bmatrix}
        \mathcal{A} & -i\mathcal{B} \\
        -i\mathcal{B}^* & \mathcal{A} \\
    \end{bmatrix}, \qquad \mathcal{A} = |\mathcal{Z}|^2 - \mathcal{W}^2, \quad \mathcal{B} = 2\mathcal{Z}\mathcal{W},
\end{equation}
we obtain the eigenvalues
\begin{equation}
    e^{2i\chi_\pm} = \mathcal{A} \pm i |\mathcal{B}|.
\end{equation}
From this we compute
\begin{equation}
    e^{i\chi_\pm} = \pm \sqrt{\frac{1}{2}\left( \mathbb{I}_{\scriptscriptstyle N} + \mathcal{A} \right)} + i \sqrt{\frac{1}{2}\left( \mathbb{I}_{\scriptscriptstyle N} - \mathcal{A} \right)} = \pm |\mathcal{Z}| + i \mathcal{W},
\end{equation}
taking into account that for ABSs, the matrix $\textrm{Im}(e^{i\chi_\pm})$ is positive semi-definite. Energies are proportional to the real part of $e^{i\chi_\pm}$, i.e. $E_\pm = \pm |\Delta \mathcal{Z}| = \pm |\Delta| \sqrt{\mathbb{I}_{\scriptscriptstyle N} - \mathcal{T} \sin(\phi/2)^2}$. The corresponding (electron part) eigenfunctions are
\begin{equation}
    \psi'_{e,\pm} = \frac{1}{\sqrt{2}} \begin{bmatrix}
        \mathcal{Z}/|\mathcal{Z}| \\
        \mp \mathbb{I}_{\scriptscriptstyle N}
    \end{bmatrix}
\end{equation}
from which the hole part is obtained by $ \psi'_{h,\pm} = e^{-i\chi_\pm} (S')^* \psi'_{e,\pm} $, giving the complete set of ABSs:
\begin{equation}
    \psi'_\pm = \frac{1}{\sqrt{2}} \begin{bmatrix}
        \psi'_{e,\pm} \\
        \psi'_{h,\pm}
    \end{bmatrix} = \frac{1}{2}\begin{bmatrix}
        \mathcal{Z}/|\mathcal{Z}|\\
        \mp \mathbb{I}_{\scriptscriptstyle N}  \\
        \pm \mathbb{I}_{\scriptscriptstyle N} \\
        \mathcal{Z}/|\mathcal{Z}|\\
    \end{bmatrix}.
\end{equation}
This can be compactly written in matrix form as
\begin{equation}
    \begin{bmatrix}
        \psi'_+ & \psi'_-
    \end{bmatrix} = \begin{bmatrix}
        \psi'_a & \psi'_b
    \end{bmatrix} \begin{bmatrix}
        \mathcal{Z} / |\mathcal{Z}| & 0 \\
        0 & \mathbb{I}_{\scriptscriptstyle N}
    \end{bmatrix} \mathcal{H}_{\scriptscriptstyle 2N} , \qquad \begin{bmatrix}
        \psi'_a & \psi'_b
    \end{bmatrix} = \frac{1}{\sqrt{2}} \begin{bmatrix}[r]
        \mathbb{I}_{\scriptscriptstyle N} & 0 \\
        0 & -\mathbb{I}_{\scriptscriptstyle N} \\
        0 & \mathbb{I}_{\scriptscriptstyle N} \\
        \mathbb{I}_{\scriptscriptstyle N} & 0 \\
    \end{bmatrix}.
\end{equation}
We will write the Hamiltonian projected to the subspace of ABSs in terms of the following basis
\begin{equation}
    \begin{bmatrix}
        \psi_\alpha & \psi_\beta
    \end{bmatrix} = \mathcal{U}^\dagger \begin{bmatrix}
        \psi'_a & \psi'_b
    \end{bmatrix} \mathcal{H}_{\scriptscriptstyle 2N} e^{i \hat{\phi}/2} = \frac{1}{\sqrt{2}}\begin{bmatrix}
        0 & V_1 \\
        V_2 & 0 \\
        U_1 & 0 \\
        0 & -U_2 \\
    \end{bmatrix}
\end{equation}
which is independent of the superconducting phase $\phi$. The projected Hamiltonian is then
\begin{align}
    H_{p} &= \mathcal{U}^\dagger \begin{bmatrix}
        \psi'_+ & \psi'_-
    \end{bmatrix} \begin{bmatrix}
         |\Delta \mathcal{Z}| & 0 \\
         0 & -|\Delta \mathcal{Z}|
    \end{bmatrix}
    \begin{bmatrix}
        (\psi'_+)^\dagger \\
        (\psi'_-)^\dagger \\
    \end{bmatrix} \mathcal{U} \nonumber \\
    &= |\Delta| \begin{bmatrix}
        \psi_\alpha & \psi_\beta
    \end{bmatrix} \begin{bmatrix}
        \textrm{Re}(\mathcal{Z}) & -i e^{-i\phi/2} \textrm{Im}(\mathcal{Z}) \\
        i e^{i\phi/2} \textrm{Im}(\mathcal{Z}) & -\textrm{Re}(\mathcal{Z}) \\
    \end{bmatrix} \begin{bmatrix}
        \psi_\alpha^\dagger \\
        \psi_\beta^\dagger
    \end{bmatrix} \nonumber \\
    &= |\Delta| \begin{bmatrix}
        \psi_\alpha & \psi_\beta
    \end{bmatrix} \begin{bmatrix}
        \sqrt{\mathcal{T}} \cos(\phi/2) & i e^{-i\phi/2} \sqrt{\mathcal{R}} \\
        -i e^{i\phi/2} \sqrt{\mathcal{R}} & -\sqrt{\mathcal{T}} \cos(\phi/2) \\
    \end{bmatrix} \begin{bmatrix}
        \psi_\alpha^\dagger \\
        \psi_\beta^\dagger
    \end{bmatrix} \nonumber \\
    \label{eq:H_ABS:projected}
    & = |\Theta \rangle \left( \vec{h}_{\scriptscriptstyle N} \cdot (\vec{\Sigma} \otimes \mathbb{I}_{\scriptscriptstyle N}) \right) \langle \Theta | 
\end{align}
where in the last step we defined a shortened notation $|\Theta \rangle \equiv \begin{bmatrix} \psi_\alpha & \psi_\beta \end{bmatrix}$, $\vec{\Sigma}$ denotes a vector of Pauli matrices acting in the space of states $\psi_\alpha$ and $\psi_\beta$, and we defined a $3N$ dimensional analog of a magnetic field:
\begin{equation}
    \vec{h}_{\scriptscriptstyle N} = |\Delta| \begin{bmatrix}[r]
        \sqrt{\mathcal{R}} \sin(\phi/2) \\
        -\sqrt{\mathcal{R}} \cos(\phi/2) \\
        \sqrt{\mathcal{T}} \cos(\phi/2) \\
    \end{bmatrix}.
\end{equation}
Considering the polar decomposition of the scattering matrix [Eq. (\ref{eq:polarDecomposition})], projector $P = |\Theta \rangle \langle \Theta |$ and the projected Hamiltonian $H_{p}$ can be expressed in terms of reflection and transmission matrices:
\begin{subequations}
    \begin{align}
        P &= \frac{1}{2}\begin{bmatrix}
                \mathbb{I}_{\scriptscriptstyle N} & 0 & 0 & -(t' t'^\dagger)^{-1/2} t' \\
                0 & \mathbb{I}_{\scriptscriptstyle N} & (t t^\dagger)^{-1/2} t & 0 \\
                0 & t^\dagger (t t^\dagger)^{-1/2} & \mathbb{I}_{\scriptscriptstyle N} & 0 \\
                -t'^\dagger (t' t'^\dagger)^{-1/2} & 0 & 0 & \mathbb{I}_{\scriptscriptstyle N} \\
            \end{bmatrix}, \\
        H_{p} &= \frac{1}{2}\begin{bmatrix}
            H_{e} & \mathcal{H} \\
            \mathcal{H}^\dagger & H_{h} \\
        \end{bmatrix} = P \begin{bmatrix}
            0 & \mathcal{H} \\
            \mathcal{H}^\dagger & 0 \\
        \end{bmatrix}P,
    \end{align}
\end{subequations}
where
\begin{subequations}
    \begin{align}
        H_e &= |\Delta| \begin{bmatrix}
            -\cos(\frac{\phi}{2}) (t' t'^\dagger)^{1/2} & e^{i \phi/2} r t^\dagger (t t^\dagger )^{-1/2} \\
            e^{-i \phi/2} (t t^\dagger )^{-1/2} t r^\dagger & \cos(\frac{\phi}{2})   (t t^\dagger)^{1/2} \\
        \end{bmatrix}, \\
        H_h &= S^\dagger e^{-i \hat{\phi}} H_e  e^{i \hat{\phi}} S = |\Delta|\begin{bmatrix}
            \cos(\frac{\phi}{2}) (t^\dagger t)^{1/2} & e^{-i \phi/2} t^\dagger (t t^\dagger )^{-1/2} r' \\
            e^{i \phi/2} r'^\dagger (t t^\dagger )^{-1/2} t & -\cos(\frac{\phi}{2}) (t'^\dagger t')^{1/2} \\
        \end{bmatrix}, \\
        \mathcal{H} &= \frac{|\Delta|}{2} \left\{S, \, e^{i \hat{\phi}}\right\} = |\Delta| \begin{bmatrix}
            e^{i\phi/2} r & \cos(\frac{\phi}{2}) t' \\
            \cos(\frac{\phi}{2}) t & e^{-i\phi/2} r' \\
        \end{bmatrix}.
    \end{align}
\end{subequations}
This concludes the derivation of the effective Hamiltonian for ABSs. Note that the block $H_e$ is the same as the Hamiltonian shown in Eq. (9) of the main text.

\subsection{Chiral symmetry}

The chiral symmetry (CS) [Eq. (6b) of the main text] implies that eigenfunctions come in pairs, where $\Psi(\vec{\alpha}, -\vec{\phi}) = i \tau_y \Psi(\vec{\alpha}, \vec{\phi})$. If we represent the wavefunction $\Psi(\vec{\alpha}, \vec{\phi})$ in the $m$-th lead in terms of incoming ($a_e$, $a_h$) and outgoing ($b_e$, $b_h$) waves, we have
\begin{equation}
    \Psi(\vec{\alpha}, \vec{\phi}) = \mathcal{N} \begin{bmatrix}
        e^{-ikx} a_e + e^{ikx} b_e \\
        e^{ikx} a_h + e^{-ikx} b_h
    \end{bmatrix},
\end{equation}
where $\mathcal{N}$ is normalization factor, $x$ is the distance from the normal region, and $k$ is the wave vector along $x$ in the $m$-th lead. Thus, the CS effectively changes $(b_e, b_h)$ to $(a_h, -a_e)$. This translates for the Hamiltonian projected to ABSs [Eq. (\ref{eq:H_ABS:projected})], which is written in the basis of outgoing waves (as stated in the main text), to
\begin{align}
        H_p(\vec{\alpha}, \vec{\phi}) \begin{bmatrix}
        b_e \\
        b_h
    \end{bmatrix} &= E(\vec{\alpha}, \vec{\phi}) \begin{bmatrix}
        b_e \\
        b_h
    \end{bmatrix}, \\
    H_p(\vec{\alpha}, -\vec{\phi}) \begin{bmatrix}[r]
        a_h \\
        -a_e
    \end{bmatrix} &= -E(\vec{\alpha}, \vec{\phi}) \begin{bmatrix}[r]
        a_h \\
        -a_e
    \end{bmatrix}.
\end{align}
Considering the relation between incoming and outgoing waves \citep{beenakker1991universal}
\begin{subequations}
    \begin{equation}
        \begin{bmatrix}[r]
            a_e \\
            a_h
        \end{bmatrix} = e^{-i\chi} \begin{bmatrix}
            0 & e^{i\hat{\phi}} \\
            e^{-i\hat{\phi}} & 0 \\
        \end{bmatrix} \begin{bmatrix}
            b_e \\
            b_h
        \end{bmatrix},
    \end{equation}
\end{subequations}
we get
\begin{equation}
    \begin{bmatrix}
        e^{i\hat{\phi}} & 0 \\
        0 & -e^{-i\hat{\phi}} \\
    \end{bmatrix} H_p(\vec{\alpha}, -\vec{\phi}) \begin{bmatrix}
        e^{-i\hat{\phi}} & 0 \\
        0 & -e^{i\hat{\phi}} \\
    \end{bmatrix} = -H_p(\vec{\alpha}, \vec{\phi}).
\end{equation}
In the two-terminal case, for $\phi=0$ and $\phi=\pi$ we have $\Gamma_0 = \tau_z \otimes \nu_0 \otimes \mathbb{I}_{\scriptscriptstyle N}$ and $\Gamma_\pi = \tau_z \otimes \nu_z \otimes \mathbb{I}_{\scriptscriptstyle N}$, respectively, where $\nu_0$ is $2 \times 2$ identity matrix acting in the space of left and right outgoing waves, and $\nu_z$ is the Pauli matrix acting in the same space. If we drop the particle-hole degree of freedom and restrict ourselves to only the electron part as we did in the main text, $\Gamma_\pi$ becomes $(\nu_z \otimes \mathbb{I}_{\scriptscriptstyle N})$, as presented in the main text.

Acting with $\Gamma_\pi$ on $| \Theta \rangle$, we find that $\Gamma_\pi | \Theta \rangle = -| \Theta \rangle (\Sigma_z \otimes \mathbb{I}_{\scriptscriptstyle N})$. This demonstrates that at $\phi = \pi$, the states $\psi_\alpha$ and $\psi_\beta$ are chiral with chirality $-$ and $+$, respectively. In the remainder of this text, we will omit writing $\mathbb{I}_{\scriptscriptstyle N}$ for brevity.

\subsection{Winding number}
Proceeding to the computation of the winding number at $\phi=\pi$, we express the flattened Hamiltonian as \citep{shiozaki2014topology}
\begin{equation}
    Q = |\Theta \rangle \left( \hat{h}_{\scriptscriptstyle N} \cdot \vec{\Sigma} \right) \langle \Theta |,
\end{equation}
where $\hat{h}_{\scriptscriptstyle N} = \vec{h}_{\scriptscriptstyle N} / | \vec{h}_{\scriptscriptstyle N} | = [1, 0, 0]^T$.

Let us first compute $\langle \Theta |\Theta \rangle$ and $\langle \Theta | \textrm{d} \Theta \rangle$:
\begin{subequations}
    \begin{align}
        \langle \Theta |\Theta \rangle &= \begin{bmatrix}
            \psi_\alpha^\dagger \psi_\alpha & 0 \\
            0 & \psi_\beta^\dagger \psi_\beta
        \end{bmatrix} = \begin{bmatrix}
            \mathbb{I}_{\scriptscriptstyle N} & 0 \\
            0 & \mathbb{I}_{\scriptscriptstyle N}
        \end{bmatrix},\\
        \langle \Theta | \textrm{d} \Theta \rangle &= \begin{bmatrix}
            \psi_\alpha^\dagger \textrm{d} \psi_\alpha & 0 \\
            0 & \psi_\beta^\dagger \textrm{d} \psi_\beta
        \end{bmatrix} = \frac{1}{2}\begin{bmatrix}
            V_2^\dagger \textrm{d} V_2 + U_1^\dagger \textrm{d} U_1 & 0 \\
            0 & V_1^\dagger \textrm{d} V_1 + U_2^\dagger \textrm{d} U_2
        \end{bmatrix}.
    \end{align}
\end{subequations}
Now we evaluate the winding number:
\begin{align}
    W_\pi &= \frac{1}{4\pi i} \oint_{\mathcal{C}_\alpha} \textrm{tr}_p\left[ \Gamma_\pi Q \textrm{d} Q \right] \nonumber \\
    & = \frac{1}{4\pi i} \oint_{\mathcal{C}_\alpha} \textrm{tr}_p\left[ |\Theta \rangle (-\Sigma_z) \Sigma_x \langle \Theta | \textrm{d} Q \right] \nonumber \\
    & = \frac{1}{4\pi i} \oint_{\mathcal{C}_\alpha} \textrm{tr}\left[ -\Sigma_y \langle \Theta | \textrm{d} \left( |\Theta \rangle \Sigma_x \langle \Theta | \right) |\Theta \rangle \right] \nonumber \\
    & = \frac{1}{2\pi i} \oint_{\mathcal{C}_\alpha} \textrm{tr}\left[ -\Sigma_z \langle \Theta | \textrm{d}  |\Theta \rangle \right] \nonumber \\
    & = \frac{1}{4\pi i} \oint_{\mathcal{C}_\alpha} \textrm{tr}\left[ V_1^\dagger \textrm{d} V_1 + U_2^\dagger \textrm{d} U_2 - V_2^\dagger \textrm{d} V_2 - U_1^\dagger \textrm{d} U_1 \right],
\end{align}
where $\mathcal{C}_\alpha$ denotes a connected path in the space of AC fluxes $\vec{\alpha} = [\alpha_1, \, \alpha_2]^T$, $\textrm{tr}_p$ indicates tracing over all ABSs, and $\textrm{tr}$ is the trace of a matrix. Considering the polar decomposition of $r$ and $r'$, the above expression can be rewritten as
\begin{equation}
    W_\pi = \frac{1}{4\pi i} \oint_{\mathcal{C}_\alpha} \textrm{tr} \left[ r^{-1} \textrm{d}r - (r')^{-1}\textrm{d} r' \right] = \frac{1}{2} \left( W_r - W_{r'} \right)
\end{equation}
where we define the winding numbers $W_r$ and $W_{r'}$ for the reflection matrices $r$ and $r'$, respectively.

From $\textrm{det}\left( r t^\dagger r' t'^\dagger \right) = - \textrm{det}\left( \mathcal{R} \mathcal{T} \right)$, we derive the relation
\begin{equation}
    W_r + W_{r'} = W_t + W_{t'},
\end{equation}
where in the derivation, we assume that $\mathcal{R}$ and $\mathcal{T}$ are not singular anywhere on the path $\mathcal{C}_\alpha$, ensuring that derivatives and winding numbers are well-defined. If the path $\mathcal{C}_\alpha$ does not enclose any singularities in $t$ or $t'$, i.e. no gap touching singularities, then $W_r + W_{r'} = W_t + W_{t'} = 0$. Thus, we obtain the result for the winding number at $\phi = \pi$:
\begin{equation}
    W_\pi = W_r = \frac{1}{2\pi i} \oint_{\mathcal{C}_\alpha} \textrm{tr} \left[ r^{-1} \textrm{d}r \right] = \frac{1}{2\pi i} \oint_{\mathcal{C}_\alpha} \textrm{d} \, \log \left[ \textrm{det} \left( r \right) \right]
\end{equation}
which is the same as the one presented in the main text under the same assumptions.

\subsection{Low-energy Hamiltonian}

Let us derive the low-energy Hamiltonian near the Weyl node at $\vec{x}_{\scriptscriptstyle W}$. At $\vec{x}_{\scriptscriptstyle W}$, we choose for the basis the two chiral states crossing at the Weyl node (indicated by subscript $0$), denoted as $|a^-\rangle \equiv |\psi_{\alpha,0}(\vec{x}_{\scriptscriptstyle W}) \rangle$ and $|a^+\rangle  \equiv |\psi_{\beta,0}(\vec{x}_{\scriptscriptstyle W}) \rangle$ (note the computation of chiralities of $\psi_{\alpha,\beta}$ at the end of the previous subsection). Additionally, $H_p(\vec{x}_{\scriptscriptstyle W})|a^\pm\rangle = 0$. Expressing the eigenfunctions at $\vec{x}$, where $|\vec{x}- \vec{x}_{\scriptscriptstyle W}| \ll 1$, in terms of this basis, we have:
\begin{equation}
    | \psi_{0}^\pm (\vec{x})\rangle \approx \alpha^\pm |a^+\rangle + \beta^\pm |a^-\rangle.
\end{equation}
From here, we obtain
\begin{equation}
    H_p(\vec{x}) | \psi_{0}^\pm (\vec{x})\rangle \approx \left(H_p(\vec{x}_{\scriptscriptstyle W}) + \delta\vec{x} \cdot \nabla_{\vec{x}} H_p(\vec{x}_{\scriptscriptstyle W}) \right) \left( \alpha^\pm |a^+\rangle + \beta^\pm |a^-\rangle \right) = E_0^\pm(\vec{x}) \left( \alpha^\pm |a^+\rangle + \beta^\pm |a^-\rangle \right)
\end{equation}
where $\delta\vec{x} = \vec{x}- \vec{x}_{\scriptscriptstyle W}$. Projecting this onto the states $|a^\pm\rangle$, we obtain a linear system:
\begin{equation}
    \begin{bmatrix}
        \langle a^+ | \\
        \langle a^- | 
    \end{bmatrix} \left( \delta\vec{x} \cdot \nabla_{\vec{x}} H_p(\vec{x}_{\scriptscriptstyle W}) \right) \begin{bmatrix}
        | a^+ \rangle & | a^- \rangle 
    \end{bmatrix} \begin{bmatrix}
        \alpha^\pm \\
        \beta^\pm
    \end{bmatrix} = E_0^\pm(\vec{x}) \begin{bmatrix}
        \alpha^\pm \\
        \beta^\pm
    \end{bmatrix}.
\end{equation}
From this, we can derive the Weyl Hamiltonian in the basis of chiral states $| a^\pm \rangle$:
\begin{equation}
    H_{\scriptscriptstyle W} = \delta\vec{x} \cdot M_3 \vec{\Sigma}
\end{equation}
where
\begin{equation}
    M_3 \vec{\Sigma} = \begin{bmatrix}
        \langle a^+ | \\
        \langle a^- | 
    \end{bmatrix} \left( \nabla_{\vec{x}} H_p(\vec{x}_{\scriptscriptstyle W}) \right) \begin{bmatrix}
        | a^+ \rangle & | a^- \rangle 
    \end{bmatrix}.
\end{equation}
Now, let us evaluate the matrix $M_3$ row by row. Starting with $\alpha_i$, we have:
\begin{subequations}
    \begin{align}
        \label{eq:M3alpha:step1}
        (M_3)_{[\alpha_i\bullet]} \vec{\Sigma} &= \begin{bmatrix}
            \langle a^+ | \\
            \langle a^- | 
        \end{bmatrix} \left.\left( \partial_{\alpha_i} H_p(\vec{x}) \right)\right|_{\vec{x} = \vec{x}_{\scriptscriptstyle W}} \begin{bmatrix}
            | a^+ \rangle & | a^- \rangle 
        \end{bmatrix} \\
        \label{eq:M3alpha:step2}
        &= \left. \partial_{\alpha_i} \left( \begin{bmatrix}
            \langle a^+ | \\
            \langle a^- | 
        \end{bmatrix} | \Theta(\vec{\alpha}) \rangle \left( \Sigma_x \otimes \sqrt{\mathcal{R}(\vec{\alpha})} \right) \langle \Theta(\vec{\alpha}) | \begin{bmatrix}
            | a^+ \rangle & | a^- \rangle 
        \end{bmatrix} \right) \right|_{\vec{\alpha} = \vec{\alpha}_{\scriptscriptstyle W}} \\
        &\approx \left. \partial_{\alpha_i} \left( \begin{bmatrix}
            0 & \langle a^+ | \psi_{\beta,0} (\vec{\alpha}) \rangle \\
            \langle a^- | \psi_{\alpha,0} (\vec{\alpha}) \rangle & 0 \\
        \end{bmatrix} \begin{bmatrix} 
            0 & \sqrt{\mathcal{R}_0(\vec{\alpha})} \\
            \sqrt{\mathcal{R}_0(\vec{\alpha})} & 0 \\
        \end{bmatrix} \begin{bmatrix}
            0 & \langle \psi_{\alpha,0} (\vec{\alpha}) | a^- \rangle \\
            \langle \psi_{\beta,0} (\vec{\alpha}) | a^+ \rangle & 0 \\
        \end{bmatrix} \right) \right|_{\vec{\alpha} = \vec{\alpha}_{\scriptscriptstyle W}} \\
        &= \begin{bmatrix}
            0 & \partial_{\alpha_i} \tilde{\zeta}(\vec{\alpha}_{\scriptscriptstyle W}) \\
            \partial_{\alpha_i} \tilde{\zeta}^*(\vec{\alpha}_{\scriptscriptstyle W}) & 0 \\
        \end{bmatrix} = {\rm Re}(\partial_{\alpha_i} \tilde{\zeta}(\vec{\alpha}_{\scriptscriptstyle W})) \Sigma_x - {\rm Im}(\partial_{\alpha_i} \tilde{\zeta}(\vec{\alpha}_{\scriptscriptstyle W})) \Sigma_y,
    \end{align}
\end{subequations}
where $\mathcal{R}_0$ is the smallest singular value of $r r^\dagger$, and $\tilde{\zeta}(\vec{\alpha}) =  \langle a^+ | \psi_{\beta,0} (\vec{\alpha}) \rangle \sqrt{\mathcal{R}_0(\vec{\alpha})} \langle \psi_{\alpha,0} (\vec{\alpha}) | a^- \rangle$ is a complex scalar.
Similarly, for $\phi$, we have:
\begin{align}
    (M_3)_{[\phi\bullet]} \vec{\Sigma} &= \begin{bmatrix}
        \langle a^+ | \\
        \langle a^- | 
    \end{bmatrix} \left.\left( \partial_{\phi} H_p(\vec{x}) \right)\right|_{\vec{x} = \vec{x}_{\scriptscriptstyle W}} \begin{bmatrix}
        | a^+ \rangle & | a^- \rangle 
    \end{bmatrix} \nonumber \\
    &\approx \!\! \left. \partial_{\phi} \!\!\left( \!\begin{bmatrix}
        0 & \langle a^+ | \psi_{\beta,0} (\vec{\alpha}_{\scriptscriptstyle W}) \rangle \\
        \langle a^- | \psi_{\alpha,0} (\vec{\alpha}_{\scriptscriptstyle W}) \rangle & 0 \\
    \end{bmatrix} \!\!\! \begin{bmatrix} 
        \cos(\phi/2) & 0 \\
        0 & -\cos(\phi/2) \\
    \end{bmatrix} \!\!\! \begin{bmatrix}
        0 & \langle \psi_{\alpha,0} (\vec{\alpha}_{\scriptscriptstyle W}) | a^- \rangle \\
        \langle \psi_{\beta,0} (\vec{\alpha}_{\scriptscriptstyle W}) | a^+ \rangle & 0 \\
    \end{bmatrix} \! \right) \!\right|_{\phi = \pi} \nonumber \\
    &= \Sigma_x \left(-\frac{1}{2}\Sigma_z\right) \Sigma_x = \frac{1}{2} \Sigma_z.
\end{align}
Thus, the matrix $M_3$ can be written in the same form as in the main text
\begin{equation}
    M_3 \!=\! 
    \left[ \begin{array}{ *{4}{c} }
    & & 0 \\
    \multicolumn{2}{c}
      {\raisebox{\dimexpr\normalbaselineskip+.7\ht\strutbox-1.2\height}[0pt][0pt]{\scalebox{1.5}{$M_2$}}} & 0 \\
    0 & 0 & \frac{1}{2}
    \end{array} \right]\!,\qquad M_2 \!=\! \begin{bmatrix}
    \partial_{\alpha_1} {\rm Re}(\tilde{\zeta}) & -\partial_{\alpha_1} {\rm Im}(\tilde{\zeta}) \\
    \partial_{\alpha_2} {\rm Re}(\tilde{\zeta}) & -\partial_{\alpha_2} {\rm Im}(\tilde{\zeta}) \end{bmatrix}_{\vec{\alpha} = \vec{\alpha}_{\scriptscriptstyle W}},
\end{equation}
although with $\tilde{\zeta}$ replacing $\zeta$ of the main text. Still, the same argument that $W_{\scriptscriptstyle W} = \textrm{sgn} \left(\textrm{det}(M_2) \right) = \textrm{sgn}  \left(\textrm{det}(M_3) \right) = C_{\scriptscriptstyle W}$ can be made. This also demonstrates the dispersion is conical at $\vec{x}_{\scriptscriptstyle W}$ with the low-energy dispersion being
\begin{equation}
    E_0^\pm(\delta\vec{x}) = \pm \sqrt{\left|\delta\vec{\alpha} \cdot \nabla_{\vec{\alpha}} \tilde{\zeta}\right|^2 + \left(\frac{\delta \phi}{2}\right)^2}.
\end{equation}

Let us now demonstrate the connection between $\tilde{\zeta}$ and $\zeta$ of the main text. Equation (\ref{eq:M3alpha:step1}) can be expressed as
\begin{align}
    \left. \partial_{\alpha_i} \left( \frac{1}{2} \begin{bmatrix}
        \langle a_e^+ | & \langle a_h^+ | \\
        \langle a_e^- | & \langle a_h^- | 
    \end{bmatrix} H_p(\vec{\alpha}, \phi=\pi) \begin{bmatrix}
        | a_e^+ \rangle & | a_e^- \rangle \\
        | a_h^+ \rangle & | a_h^- \rangle 
    \end{bmatrix} \right) \right|_{\vec{\alpha} = \vec{\alpha}_{\scriptscriptstyle W}}
\end{align}
with $| a_{e,h}^\pm \rangle$ being the electron and hole parts of $| a^\pm \rangle$, which leads to an alternative expression for $\tilde{\zeta}$:
\begin{equation}
    \tilde{\zeta} = \frac{1}{4} \left( \langle a_e^+ | H_e | a_e^- \rangle + \langle a_e^+ | \mathcal{H} | a_h^- \rangle + \langle a_h^+ | \mathcal{H}^\dagger | a_e^- \rangle + \langle a_h^+ |  H_h | a_h^- \rangle\right),
\end{equation}
where the first (electronic) term is the same as in the main text, being equal to $\zeta$. However, as we have shown in the previous subsection, working with the full states or just with their electronic parts yields the same topological charges. Hence, while performing the full computation involving both electron and hole parts introduces additional terms in $\tilde{\zeta}$ compared to the approach outlined in the main text, this adjustment does not alter any of the conclusions regarding the system's topology.

% \bibliographystyle{ieeetr}
%\bibliographystyle{apsrev4-1}
% \bibliography{bibliography_list}